\documentclass[a4paper,11pt]{article}
\usepackage{pos}
\usepackage{subcaption}

\title{Moderator Modeling for High Intensity Slow Positron Sources}

\author*[a]{Sophie Crisp}
\author[a,b]{Ryland Goldman}
\author[a,c]{Spencer Gessner}

\affiliation[a]{SLAC National Accelerator Laboratory,\\
  2575 Sand Hill Road, Menlo Park, CA, 94025, USA}

\affiliation[b]{Department of Physics and Astronomy, University of California, Los Angeles,\\
475 Portola Plaza, Los Angeles, CA, 90095, USA}

\affiliation[c]{Department of Physics, Stanford University,\\
382 Via Pueblo Mall, Stanford, CA, 94305, USA}

\emailAdd{scrisp11@slac.stanford.edu}

\abstract{Slow positron beams enable diverse applications, from surface-sensitive materials studies to positronium physics, but progress is limited by source intensity and brightness. In linac-based sources, mono-energetic, slow positrons are produced by moderating the broad, divergent distribution of fast positrons produced by high-energy electrons incident on a high-Z target. This process is intrinsically inefficient. Conventional linac-based designs place the moderator close to the target. Tungsten moderators become less efficient when heated by high-power, fast positrons, leading to defect-related losses. Cryogenic moderators, like solid neon, melt when exposed to high-power, fast positrons. Following the method of O'Rourke et al.~\cite{orourke_2011_simulations}, we use Monte Carlo simulations combined with a diffusion model to investigate how moderator geometry, material, and incident fast positron energy affect slow positron production. We compare single tungsten foil and multi-foil configurations with solid neon moderators in reflection and transmission geometries. We find that for fast positron energies below 300 keV, solid neon offers order-of-magnitude higher efficiency than tungsten due to its larger diffusion length, whereas at MeV-scale energies multi-foil tungsten has higher efficiency due to the maximization of scattering. We also find that including a target proxy in simulation increases low-energy efficiency up to fourfold by allowing a fraction of initially reflected positrons to return to the moderator. We conclude that substantial gains in start-to-end efficiency at linac sources will likely require combining a decelerating cavity with a decoupled cryogenic moderator.}

\FullConference{International Workshop on Low Energy Electron Positron Physics at Jefferson Lab (LEEPP2026)\\
23--27 March 2026\\
Thomas Jefferson National Accelerator Facility, Virginia, USA\\}

\begin{document}
\maketitle

\section{Introduction}
Since positrons were first generated in the laboratory, there has been interest in using them not just in the high energy community for collider physics, but also in the low energy, ``slow positron" regime~\cite{coleman2000positronBook}. These slow positrons are generally defined as having up to 10 keV energy, and have been used to to measure defects in materials, conduct highly surface sensitive experiments~\cite{SchultzLynn1988_InteractionsPosSurfacesThinFilms}, as well as investigate positronium and other basic science~\cite{Cassidy2018_Positronium}. However, progress in positron based science has been limited by the availability of high-intensity, bright sources. Radioisotope sources are limited to intensities of $5\times10^5$ slow $e^+/$s~\cite{YU2026_pulsedSlowNa22}, and high-intensity, linac-based sources are limited by yields and the low-efficiency moderation process~\cite{charlton_GBAROverview}.

In the case of linac-based positron sources, the ability to deliver, high-intensity beams is limited by the high angular divergence of the fast positrons from the target as well as their broad energy distribution which spans from zero to the energy of the input electrons. Moderators are used to produce near mono-energetic, slow positrons. In the moderation process, positrons enter a material and lose almost all of their kinetic energy via scattering. Some positrons thermally diffuse to the surface where they have a probability of being re-emitted with energy equal to the work function of the material. In general, the process of moderation is highly inefficient, on the order of $10^{-4}$ slow $e^+$ per fast $e^+$~\cite{orourke_2011_simulations}.

\begin{figure}
    \centering
    \includegraphics[width=0.8\linewidth]{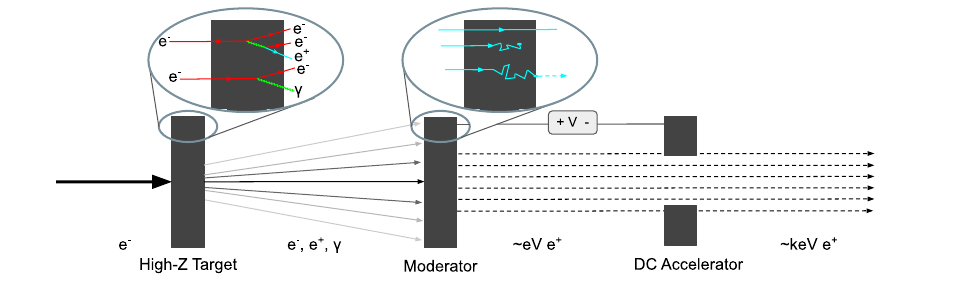}
    \caption{Schematic of a linac-based, slow positron beamline. An electron beam is incident on a high-Z target. The resulting shower of particles is incident on a moderator. Some of the positrons scatter and thermalize near the surface of the moderator, which reemits a small percentage of the incident fast positrons as near mono-energetic, slow positrons with small angular spread. These positrons are then accelerated to keV energies  and transported to experiments or remoderated for improved brightness.}
    \label{fig:basicSchematic}
\end{figure}

Slow positron sources based on linac electron sources conventionally follow the layout shown in Fig.~\ref{fig:basicSchematic}. An electron beam is incident on a high-Z target, such as tungsten, out of which a shower of positrons, electrons, and gamma rays are incident on a positron moderator, from which slow positrons are extracted. In these sources, the target and moderator are located close together with mm-scale separation. This can lead to issues with radiative heating, wherein the heating of the target transfers to the moderator, increasing the prevalence of positronium formation on the moderator surface and thereby decreasing its efficiency~\cite{charlton_GBAROverview}. In addition, the proximity can introduce defects into the moderator which also contribute to a decline in efficiency~\cite{Suzuki_1998_tungstenDefects}. Moderation efficiency is also highly energy dependent, which has led to simulation studies that explore the possibility of decelerating the fast positrons before they reach the moderator~\cite{Long_2007_decelLinac,crisp:napac2025-decelLinac}. In addition to fast positron deceleration, moving the moderator downstream of the target could enable the use of cryogenically cooled moderators. Solid neon moderators used in radioisotope have up to 1\% efficiency~\cite{mills1986SolidNeon}. For all of these reasons, physically decoupling the moderator from the target is highly desirable.

Prior work by O'Rourke et al.~\cite{orourke_2011_simulations} used the Penelope2008~\cite{penelope2008} physics model to perform Monte Carlo simulations to understand slow positron production in linac sources. In this paper, we follow the same methodology and explore a variety of geometries and materials in order to understand how these factors impact slow positron production efficiency.




\section{Simulation Setup}
\begin{figure}
    \centering
    \includegraphics[width=0.9\linewidth]{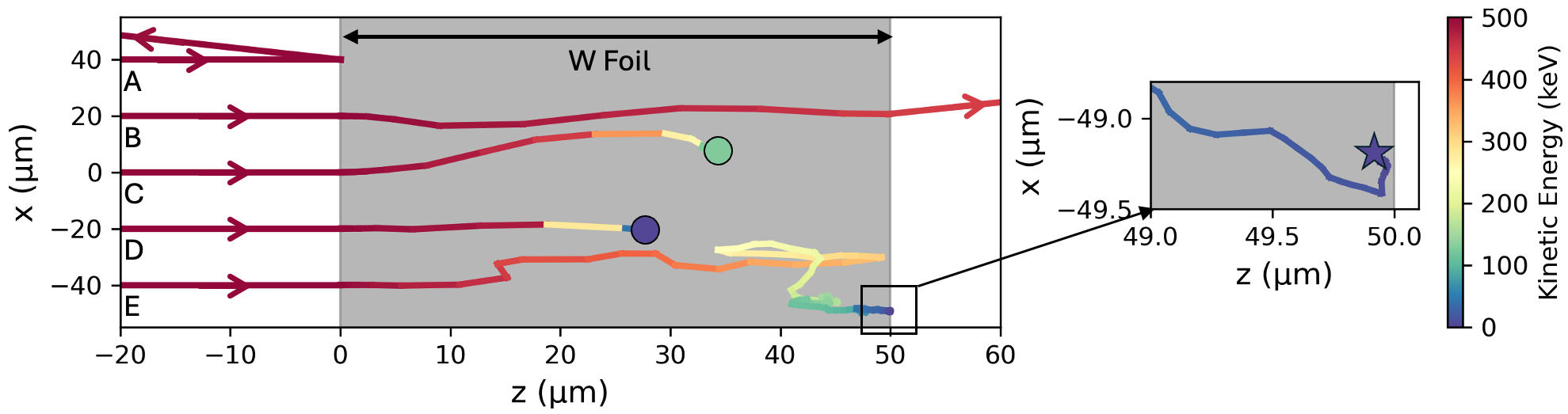}
    \caption{Pictured are the trajectories of 5 different positrons simulated inside a tungsten foil moderator, colored according to the kinetic energy of the particle along the track. The circles and star denote the locations at which the positrons annihilate in simulation. The trajectory labeled A corresponds to a positron which loses minimal energy before being scattered off in reflection. Trajectory B corresponds to a positron which scatters in the moderator but eventually is transmitted. Trajectory C corresponds to a positron which enters the moderator, but annihilates before thermalization, as seen by the high kinetic energy at its final location.  Trajectories D and E correspond to positrons which scatter and subsequently thermalize inside the moderator; these positrons are referred to as stopped. The inset shows that trajectory E corresponds to a positron that is stopped within a diffusion length of the moderator surface and therefore has a high probability of reemission.}
    \label{fig:trajectories}
\end{figure}

We use the simulation code G4beamline based on the Geant4 toolkit~\cite{agostinelli2003geant4} with the Penelope physics model combined with a diffusion model to understand the relationship between input positron energy and the probability of reemission. Positrons are tracked inside the moderator material until they reach 50~eV energy, at which point they are considered to be stopped. In Fig.~\ref{fig:trajectories}, we show the results of such tracking, highlighting 5 different possible trajectories inside of a 50~$\mu$m Tungsten foil, colored according to energy. Trajectory A corresponds to reflected positrons, B to transmission, C to annihilation, and D and E to stopped. Transmitted and reflected positrons remain at relatively high energies and are therefore not considered moderated. Trajectory C is a positron which annihilates at high energy and is therefore not able to thermally diffuse through the moderator. Only the trajectories corresponding to stopped positrons (D and E) can possibly contribute to the final moderated beam.

We assign a probability, $e^{-z/L_+}$, of diffusion to the moderator surface. The diffusion length, $L_+$, is material dependent; polycrystalline tungsten is known to have a diffusion length of $L_+=55$~nm, whereas solid neon does not have a well known diffusion length, with estimates ranging from 0.5 to 1~$\mu$m. The inset of Fig.~\ref{fig:trajectories} shows the final micron of trajectory E, positron which is stopped within a diffusion length of the surface, and therefore has a large probability of thermally diffusing to the surface. In order to match our simulation results to the efficiencies measured from radioisotope sources with cryogenic moderators, we use the $L_+=1~\mu$m for neon. Finally, the positrons which reach the moderator surface have some probability, call the branching ratio $B_R$, of being reemitted as slow positrons. For tungsten, this branching ratio is approximately $0.3$~\cite{Suzuki_1998_tungstenDefects}.

\begin{figure}
    \centering
    \includegraphics[width=0.8\linewidth]{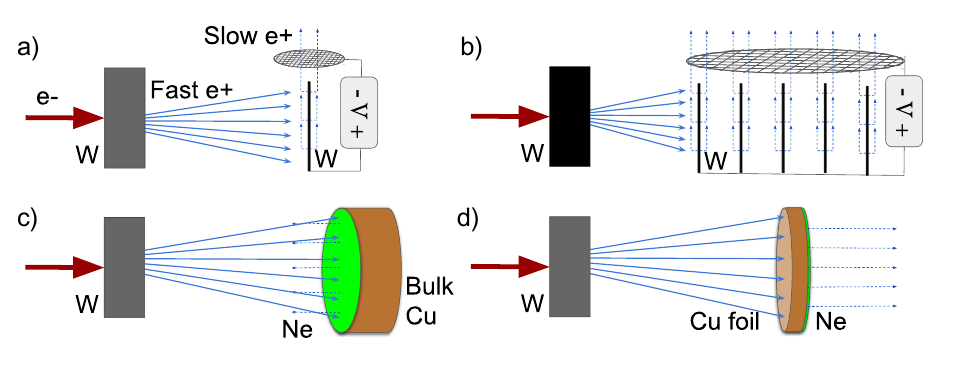}
    \caption{A high energy electron beam is incident upon a bulk tungsten target, generating high energy positrons. These fast positrons are incident on a moderator of various geometries. a) 50~$\mu$m-thick tungsten foil. b) five 50~micron-thick tungsten foils, 2 cm downstream of a tungsten block, spaced by 5~mm. In cases a and b slow positrons may be extracted transversely using a voltage differential to a grid. c) 150~$\mu$m solid neon deposited on bulk copper, moderated positrons are extracted in reflection. d) 150~$\mu$m-thick neon deposited on 25~$\mu$m-thick copper foil, moderated positrons are extracted in transmission.}
    \label{fig:geometry}
\end{figure}

Using this framework, we model several moderator geometry and material combinations, detailed in Fig.~\ref{fig:geometry}. Case (a) is a baseline geometry; we model a single 50~$\mu$m-thick tungsten foil, assuming that slow positrons could be extracted from either of the two faces. In this case, positrons are initialized directly before the moderator and the simulation volume does not contain a target, indicated in the Fig.~\ref{fig:geometry} by the grayed out tungsten block. This means that any positrons which are initially reflected off of the moderator are not able to be captured. In case (b), we add both a bulk tungsten block to act as a proxy for a target as well as addition tungsten foils in a vane configuration. The first tungsten foil is located 2~cm downstream of the tungsten target, and the subsequent foils are separated by 5~mm. Fast positrons are initialized directly before the first tungsten foil, but after the bulk tungsten block.

For solid neon moderators, the neon is deposited onto a cryogenically cooled metal. For these simulations, we choose copper as the metal substrate and simulate two possible solid neon configurations which produce slow positrons either in reflection, shown in Fig.~\ref{fig:geometry}(c), or transmission as in Fig.~\ref{fig:geometry}(d). In both cases the neon is modeled with a thickness of 150~$\mu$m. In case (d), the copper is 25~$\mu$m thick. To make use of a cryogenically cooled moderator, any proposed geometry for the target must be both a large distance and off-axis from the moderation. Therefore, the simulation contains only the neon/copper moderator assemblies, and not a bulk tungsten block.

\section{Results}
\begin{figure}
    \centering
    \includegraphics[width=\linewidth]{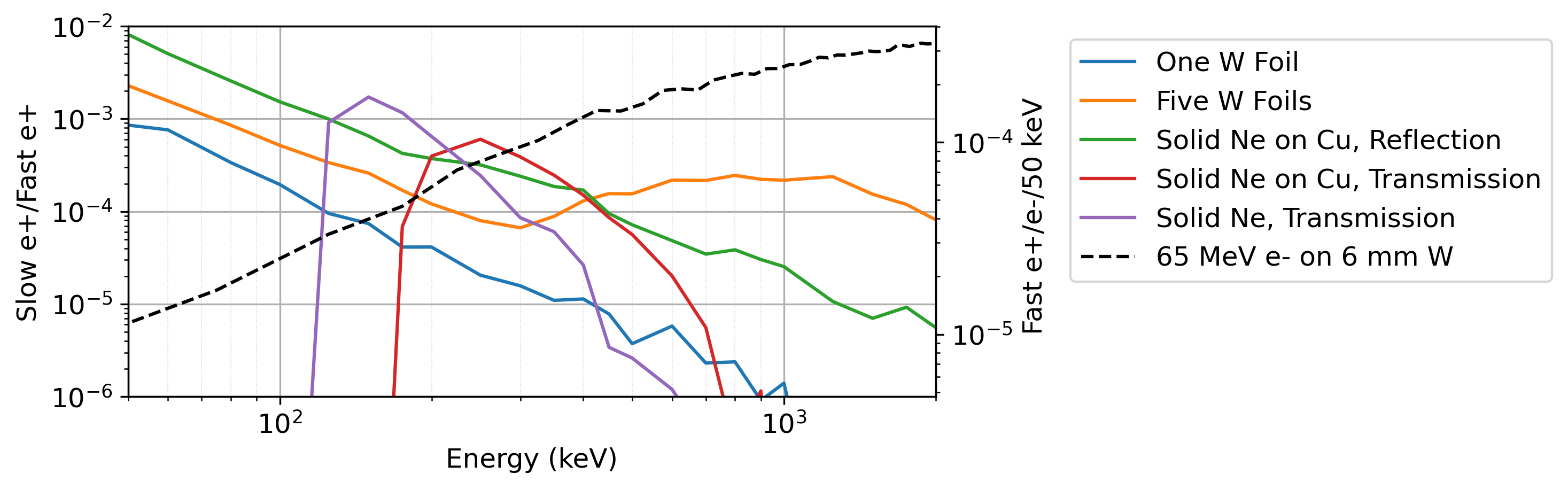}
    \caption{On the left axis and plotted with solid lines, moderation efficiency as a function of incident fast positron energy for different geometries and materials. On the right axis and plotted with a dashed line, the energy spectrum of fast positrons extracted from a simulation of 65~MeV electrons on a 6~mm Tungsten target.}
    \label{fig:ModEff}
\end{figure}

In Fig.~\ref{fig:ModEff} we show the moderation efficiency for each of the four geometries of Fig.~\ref{fig:geometry}, as well as a non-physical solid neon geometry with no copper substrate. For incident positrons with energies less than 300~keV, solid neon provides significantly higher slow positron efficiency, regardless of geometry. This is expected from theory, since the diffusion length of positrons in solid neon is at least an order of magnitude larger than that of polycrystalline tungsten, and also in practice, given that Na-22 sources emit positrons with an average energy of 215~keV. Solid neon is the preferred moderator in this case. 

Although positron sources using traps have been used to extract slow positrons in reflection~\cite{Michishio_SiCRemoderator_2022}, magnetic designs in reflection for the primary moderator can be complex~\cite{Gehrmann2023_GBarNeon}.
It would therefore be advantageous to consider the solid neon moderator in transmission, where it would both be easier to collect, confine, and transfer the initial fast positrons to the moderator, as well as allow for greater control over the final slow positron spot size and therefore beam brightness. Golge et al.~\cite{golge2014_JLABNeonMagnet} provide one such magnetic design which would allow for a high-flux primary beam to avoid contact with the transmission neon moderator via a curved solenoidal channel whose field is abruptly terminated at the moderator. We plot the efficiency of this type of solid neon transmission geometry, which peaks at an incident energy around $250$~keV and a value of $6\times10^{-4}$, significantly below the efficiencies quoted for radioisotope cryogenic moderators. Changing the thickness of the copper substrate allows for some improvement in peak efficiency -- in purple we plot the theoretical efficiency without a copper substrate, which has a peak at about $2\times10^{-3}$. Reducing the Neon thickness likewise allows for higher efficiencies, but at lower incident positron energies, which is not advantageous for linac sources.

At all energies, the single tungsten foil is comparatively inefficient. It is useful, however, to compare the single foil efficiency to the 5-foil efficiency at energies below 200~keV, where the positrons are largely unable to be transmitted through a single 50~$\mu$m-thick tungsten foil. Therefore, the only difference between the single foil and multi-foil simulation in this regime is the inclusion of the tungsten block target proxy before the moderator. For these, the inclusion of the target immediately behind the initialized positrons amounts to between a 2-4 times increase in moderator efficiency. We conclude from this that a significant fraction of moderated positrons in current linac sources are captured as a result of the close proximity between the target and moderator which allows for initially reflected particles to return to the moderator.

Meanwhile, at the MeV energy scale, the 5-foil tungsten moderator is significantly higher in efficiency than even the reflective neon geometry. This is relevant to the energy scale of linac-based positron sources. The multiple foils allow for maximal positron scattering to occur, while including enough surfaces to enable positrons to escape the moderator. Maximizing efficiency at high inicident energy, therefore, is a matter of maximizing scattering while still allowing for capture, as evidenced by the grid style moderators in use at AIST~\cite{Michishio_2025_AIST} and KEK~\cite{Wada2012}.

\section{Conclusion}
As linac-driven sources seek to increase slow positron beam brightness, it becomes increasingly necessary to decouple the moderator from the target, allowing for the use of cryogenic moderators which are known to have higher efficiencies compared to tungsten moderators. In this paper, we modeled several moderator geometries in order to understand the energy and material dependence of moderator efficiency. We found that the target is a crucial component of the total system efficiency. The proximity of the target to the moderator allows for initially reflected positrons to return to the moderator, enabling up to four times higher slow positron efficiency at low energies compared with simulations excluding the target. Our simulations also show that at low energies, the high positron diffusion length of solid neon allows for an order of magnitude improvement in moderator efficiency compared to tungsten foils, whereas at MeV scale energies the scattering enabled by multiple foils greatly increases the efficiency. This is particularly important for linac-based sources, where most positrons have energies greater than 1~MeV. In order to improve upon the overall efficiency of positron production, it may be necessary to implement both a decelerating RF cavity and cryogenic moderator scheme. Future work should model the impact of such a scheme on final slow positron beam brightness and obtain experimental data on cryogenic moderators in high-energy contexts for validation.

\acknowledgments
This work was supported by the Department of Energy, Laboratory Directed Research and Development program at SLAC National Accelerator Laboratory, under contract DE-AC02-76SF00515.
\bibliographystyle{JHEP}
\bibliography{bibl}


\end{document}